\documentclass[letterpaper, 10 pt, conference]{ieeeconf}  

\IEEEoverridecommandlockouts                              
\usepackage[noadjust]{cite}
\usepackage{amsmath,amssymb,amsfonts}
\usepackage{algorithmic}
\usepackage{graphicx}
\usepackage{textcomp}
\usepackage{xcolor}
\usepackage{float}
\usepackage{caption}
\usepackage{subcaption}
\usepackage{nicefrac}
\usepackage{booktabs}%
\usepackage[colorlinks=true, linkcolor=blue]{hyperref}

\newcommand{\ignore}[1]{}

\newcommand{\eop}{{\hfill $\Box$}}
\usepackage{dsfont}
\newtheorem{thm}{Theorem}
\newtheorem{lem}{Lemma}

\def\BibTeX{{\rm B\kern-.05em{\sc i\kern-.025em b}\kern-.08em
    T\kern-.1667em\lower.7ex\hbox{E}\kern-.125emX}}

\newcommand{\ba}{{\bf{a}}}

\newcommand{\bp}{\textbf{p}}

\newcommand{\ld }{ {\bar l}_d }
\newcommand{\p}{\textbf{p}}
\newcommand{\Q}{\mathbb Q}

\newcommand{\bqm}{{\bar q}_m}
\newcommand{\bqs}{{\bar q}_s}

\renewcommand{\S}{{\mathcal S}}

\newcommand{\Cmq}{ C_{M_q} }

\renewcommand{\P}{{\mathbb P}}
\newcommand{\M}{{\mathbb M}}
\newcommand{\G}{{\mathbb G}}

\renewcommand{\S}{{\mathbb S}}

\newcommand{\Mi}{{{\mathbb M}_i}}

\newcommand{\dbar}{\bar{d}}

\newcommand{\indc}[1]{\mathds{1}_{\left\{ #1 \right \}} }
\newcommand{\B}{{\mathcal B}}
\newcommand{\BR}{{\mathbb B}}

\newcommand{\sM}{{\mbox{\fontsize{4.7}{5}\selectfont{$\M$}}}}
\newcommand{\sMi}{{\mbox{\fontsize{4.7}{5}\selectfont{$\Mi$}}}}

\newcommand{\sS}{{\mbox{\fontsize{5.2}{5.2}\selectfont{$\S$}}}}

\newcommand{\x}{{\bf x}}

\newcommand{\pup}{ {\bar p}}

\newcommand{ \manu }{manufacturer }

 \newcommand{\manus}{manufacturers }                    
\begin{document}
\title{\LARGE \bf
Price equilibria with positive margins in  loyal-strategic markets   with discrete prices
}


\author{Gurkirat Wadhwa, Akansh Verma,  Veeraruna Kavitha and Priyank Sinha 
\thanks{Department of Industrial Engineering and Operations Research, Indian Institute of Technology, Bombay. Emails: \{gurkirat, akansh, vkavitha and priyank.sinha\}@iitb.ac.in}}

\maketitle
\thispagestyle{empty}
\pagestyle{empty}

\begin{abstract}

In competitive supply chains (SCs), pricing decisions are crucial, as they directly impact market share and profitability. Traditional SC models often assume continuous pricing for mathematical convenience, overlooking the practical reality of discrete price increments driven by currency constraints. Additionally, customer behavior, influenced by loyalty and strategic considerations, plays a significant role in purchasing decisions. To address these gaps, this study examines a SC model involving one supplier and two manufacturers, incorporating realistic factors such as customer demand segmentation and discrete price setting. Our analysis shows that the Nash equilibria (NE) among manufacturers are not unique, we then discuss the focal equilibrium. Our analysis also  reveals that low denomination factors can lead to instability as the corresponding game does not have NE. Numerical simulations demonstrate that even small changes in price increments significantly affect the competitive dynamics and market share distribution.

\end{abstract}

\section{INTRODUCTION}\label{sec_intro}

We analyze a supply chain (SC) consisting  of a single supplier and two manufacturers, where customer purchasing decisions are influenced by both loyalty and strategic considerations. Manufacturers have dedicated market bases, with some customers remaining loyal to  their manufacturers (see \cite{Wadhwa,Kantarelis}); these fractions depend upon the prices quoted for the end-product. A fraction of  non-loyal customers  of all the manufacturers are more strategic, consider many factors before  making  the purchasing decisions. The pricing strategies employed by SC agents (supplier setting raw material prices and manufacturers setting end-product prices) directly impact demand realization and, consequently, the profitability of both the manufacturers and the supplier.  

Loyal customers display brand or platform preferences (e.g., individuals consistently preferring Amazon over eBay), whereas strategic customers evaluate factors such as product reviews, price differences, and promotions before making a purchase. Notably, customer loyalty is price-sensitive; a sufficiently large price discrepancy between competing platforms can prompt even loyal customers to switch brands.
Our demand model captures these complexities of customer behavior, by considering a mean field game among strategic fraction of the customers. The resultant mean field equilibria along with price-sensitive loyal customer bases  determines the final market segmentation towards each of the   manufacturers.

We next study the impact of    loyal-strategic market segmentation  on  manufacturer and supplier profitability. We consider a three level game to analyze the same, with strategic customers at the lowest level, the two manufacturers   at the lower echelon and at the middle layer of the game and finally the supplier is at the top layer. In this paper, we analyze in detail the games at the bottom two levels. We provide a brief numerical study about the supplier optimization,  while a more detailed (top layer)  theoretical study  is postponed to the journal version. 

An often overlooked yet crucial aspect in SC literature is the discrete nature of  the price setting (see \cite{Rosenthal, Maskin}). In real-world markets, the prices are typically set in specific increments based on the smallest unit of currency; our study further takes into account such practical considerations.    


The first major finding of our research is that the agents at lower echelon can derive non-zero profit margins (depending upon the price quoted by the supplier), for the modified Bertrand duopoly  game at the lower echelon. 
These equilibria  are in contrast to the  NE of the classical  Bertrand duopoly game where the agents operate at marginal prices.  Such a contrast is obtained primarily due to the consideration of  three important  factors: i) a realistic discrete set of price choices;  ii) influence of  operating costs and iii) a better market demand model resulting from the choices of loyal and strategic customers (where the loyalty also shifts based on the prices set for the final product). 
 Further  one can have multiple NE depending upon the  price-denomination factor,   price quoted by the supplier  and other system parameters.  We then identify   a focal NE or Schelling point \cite{Schelling}  guided by natural interest of the  manufacturers -— intuitively, this would be the NE at which both the manufacturers derive maximum and equal revenue, and would not involve in price wars.

In the  numerical  study of the supplier game,  we notice that the supplier prefers to operate at a  price which can partially choke the  manufacturers ---  in this regime a  \manu setting  a strictly bigger price  derives negative utility and hence 
the two \manus set equal price (for the final product) at the corresponding focal NE.

We further consider a  numerical study of the set of the equilibria of the game at lower echelon. 
The 
numerical simulations 
indicate that the  price granularity significantly affects the competitive dynamics, and minor shifts in price denominations can lead to substantial variations. This indicates super sensitivity of the theory towards  the set of discrete prices under consideration and hence calls for a more robust theory. Our aim in future would be to work towards such a theory.

\subsection*{Literature Survey}
Many competitive models in
  supply chain (SC) literature, such as \cite{Wadhwa,Kantarelis,Zheng,Li}, 
  often overlook at the strategic nature of the non-loyal customers and assume  them to simply switch   to the competing manufacturer. 
  In contrast,   we model them as strategic players of a mean-field game that choose one among the two  manufacturers based   on price and the quality of service (QoS).  
  Authors in \cite{Maskin, Rosenthal} analyze dynamic Bertrand competition, where firms dynamically adjust the prices from among a discrete set of choices, These models either overlook important considerations like operating costs, or consider that the entire customer base switches its loyalty at one go. In contrast, our model reflects the more nuanced reality that
the extent of customer migration depends on the magnitude of
the price variations. 

\section{System Model}
 We consider a two-echelon supply chain (SC), with
 a single
 supplier at the upper echelon and  two manufacturers (indexed by $i$ and $j$) at the lower echelon. The customers purchase the final product from the manufacturers depending upon various factors like, price and the essentialness of the product, reputation of the manufacturer. The manufacturers obtain the required raw
 materials from the supplier and quote a price for the final product   depending upon their own
 market base, the price-sensitivity of the customers, the price quoted by the supplier,   the production cost, etc. The supplier quotes a price for the raw material,   depending upon the market potential of the two manufacturers, competition in the downstream market, procurement costs, etc. 

 In real world scenarios, the market is extremely heterogeneous, some customers exhibit extreme loyalties (could be buying from one particular source/\manu  for a long period  without exploring other options), some of them are extremely strategic (keep estimating the advantages of various choices  dynamically and  choosing the optimal choice continually), some of them can simply follow the majority or their friends, some can have random choices, etc. One of the  important aims of this paper 
is to construct a market model that can capture more realistic aspects like: i) \manus can have dedicated market bases;  ii) 
some customers can stop being loyal, and these numbers can depend upon the prices quoted; iii) some of them explore other options, and such numbers can depend upon the essentialness of the product; and iv) finally the exploring customers can be extremely strategic.  
 
 We  immediately begin with the detailed description of the demand model build upon the above features and which  represents the division/segmentation of the consumer market among the manufacturers.

 \subsection{ Market Segmentation}
\begin{figure}
\vspace{10mm}
    \centering
    \begin{minipage}{0.23\textwidth}
        \centering
        \includegraphics[width=\linewidth]{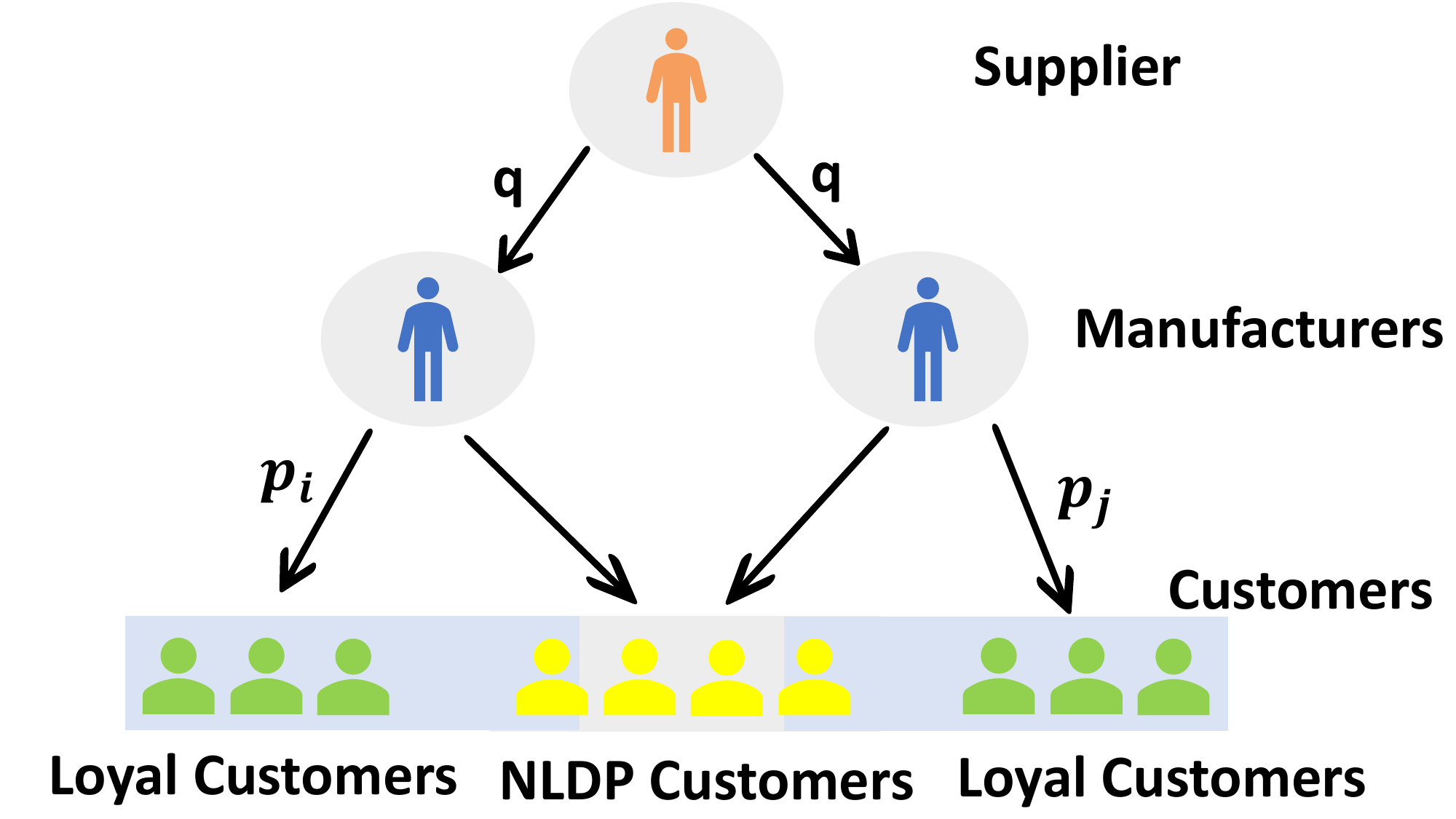}
        \label{fig:model-description}
    \end{minipage}
    \hfill
    \begin{minipage}{0.23\textwidth}
        \centering
        \includegraphics[width=\linewidth]{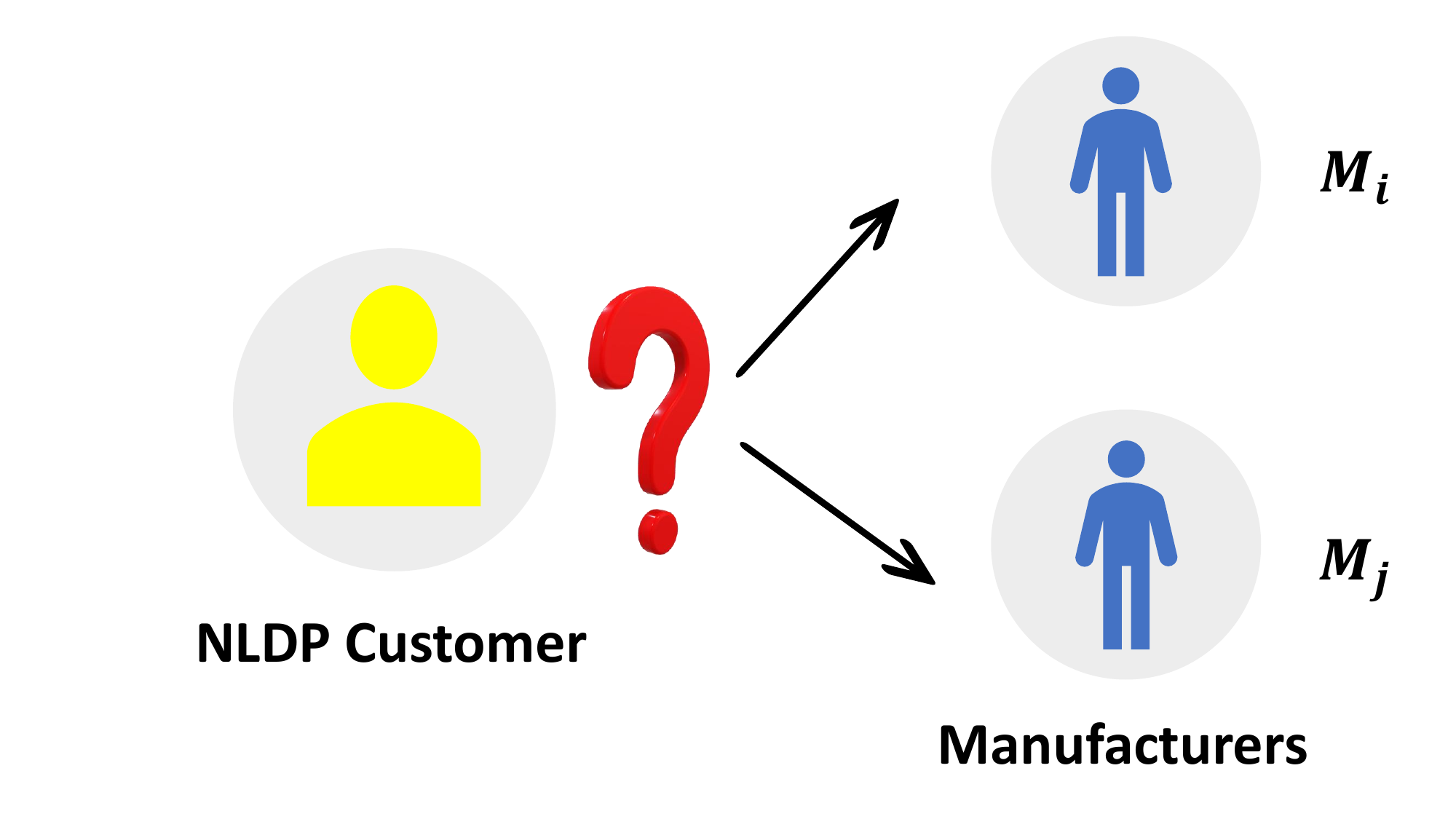}
        \label{fig:nldp}
    \end{minipage}
    \caption{Three level game and loyal-strategic customers}
    \label{fig:model}
    \vspace{-4mm}
\end{figure}

Each manufacturer (say $M_i$) has it's loyal customer base,  the maximum of this fraction or the market potential of $M_i$ is given by $\dbar$.  It  looses $\alpha p_i$  fraction of them, by quoting price $p_i$ sensitized by the parameter $\alpha$ , thus the fraction of customers that turn in at price $p_i$ is  given  by $\bar d - \alpha p_i$ (see e.g., \cite{Wadhwa}). 

Thus given  the vector of prices $(p_i, p_j)$ quoted by both the manufacturers, $\alpha (p_i +p_j)$ fraction of customers are not satisfied and do not turn into their own manufacturers. 

However, say the item is essential (and the customers are keen to buy the product),  then a fraction of them lookout for other options -- let $\varepsilon \alpha (p_i+p_j)$ represent this fraction, where $\varepsilon$ is the essentialness factor.  This fraction of customers  are Non-Loyal
    to their manufacturer (because of higher price) yet Desire the Product (hence forth referred to as NLDP customers).   These are the customers who want to buy the product but deviate from their manufacturer and look for better opportunities (see Figure \ref{fig:model}).
\begin{figure*}
\begin{eqnarray}
\Gamma = \frac{(1-\alpha\omega)}{2      \omega  }  \mbox{ and }   D_\sMi\hspace{-2mm}& \hspace{-2mm}=\hspace{-2mm}& \hspace{-2mm}\left \{
    \begin{array}{lll}     
    \left (\dbar - \alpha p_i\right )^+     & \mbox{ if }  (p_i-p_j)    \Gamma   >  2(p_i+p_j)  \alpha   \varepsilon  \\
  \left (\dbar - \alpha p_i + \varepsilon \alpha  (p_i+p_j) - (p_i - p_j) \Gamma \right )^+  
  & \mbox{ if }
   -2 (p_i+p_j)  \alpha  \varepsilon  <   (p_i-p_j)  \Gamma   < 2(p_i+p_j)  \alpha  \varepsilon    \\
      \left (\dbar - \alpha p_i + \varepsilon \alpha  (p_i+p_j) \right )^+ 
    & \mbox{ if }  (p_i-p_j) \Gamma   < - 2(p_i+p_j)  \alpha \varepsilon 
    \end{array}
    \right .  \nonumber   \\&&  \hspace{100mm} \mbox{ if }  p_j \ne n_o \nonumber \\
    D_\sMi \hspace{-2mm}& \hspace{-2mm}=\hspace{-2mm}& \hspace{-2mm} \dbar (1+ \varepsilon) - \alpha p_i  \hspace{83mm}  \mbox{ if }  p_j = n_o. \label{eqn_demand_overall}  
\end{eqnarray}
\end{figure*}
Since the NDLP customers are large in number, we require a \textit{mean field game} \footnote{\label{footnote_mean_field}Any mean field game is described by utility function $\pi (a, \mu)$, where $a \in {\cal A}$,   the set of available actions and $\mu$ a probability measure on $\cal A$ represents the empirical measure of  actions choose by the  population.  The population choice measure $\mu^*$ is the mean-field Nash equilibrium  if $\mu^{*}$ satisfies the following:
\begin{eqnarray*}
  S(\mu^{*}) &\subset& \mathcal{A}_{\mu}(\mu^{*}), \quad \text{where} \\
  S(\mu) &=& \text{support}(\mu) := \left\{ a \in \mathcal{A} : \mu_{a} > 0 \right\}, \\
  \mathcal{A}_{\mu}(\mu) &:=& \arg\max_{a\in \mathcal{A}} \pi(a,\mu).
\end{eqnarray*}} to study the  choices of this  fraction of customers at some appropriate equilibrium (see~\cite{Raghu}).  

We begin with describing the utility function of the NDLP customers. The action set of the customers is ${\cal A} = \{ M_i, M_j\}$ --- each customer has to choose one among the two manufacturers. It is obvious that  the  (negative) utility of any customer is proportional to the price set by the \manu of its choice. The   utility of these customers can further  depend upon the quality of service  (QoS) provided by the  \manu chosen;
the QoS of any system degrades as the number of customers increase, hence 
we assume the QoS $\eta$  to be a decreasing function (for example it can reflect congestion cost like delays in service, or the quality of the final product, etc).  
In all,  the  (negative) utility of each customer is defined as follows:
$$
\pi (a, \mu) =  p_a + \omega \eta (\mu_a \varepsilon\alpha (p_i+p_j)  +\dbar - \alpha p_a)  \mbox{ for all } \mu, $$
 where  $ a\in \{M_i, M_j\}
$, $\mu = (\mu_i, \mu_j)$ represents 
the division of NDLP customers across the two \manus, $\omega$ is the trade-off factor and  $\mu_a \epsilon\alpha (p_i+p_j)  +\dbar - \alpha p_a$ represents the size of the  total customer base of   \manu $a$ under NDLP-population measure $\mu$ (and at price choices $p_i, p_j$). 

We consider a linear model for QoS, $\eta(\mu) = h - \mu$ (any other  constant 
multiple can be absorbed into $\omega$). 
It is  easy to derive the mean-field equilibrium $\mu^* = (\mu_i^*, 1-\mu_i^*)$ for this game (see footnote \ref{footnote_mean_field} for definition of NE), which is given by:
\begin{equation}
\label{Eqn_NDLP_split}
\mu^*_i = \left \{
\begin{array}{ll}
 0    & \hspace{-15mm}\mbox{ if }  (p_i -p_j) (1- \alpha \omega )   >  (p_i+p_j )  \alpha \omega \varepsilon \\
 1    & \hspace{-15mm} \mbox{ if }   (p_i-p_j) (1- \alpha \omega  )    < - (p_i+p_j)  \alpha \omega \varepsilon \\  
     \frac{1}{2}-
\frac{ (p_i - p_j) (1- \alpha\omega )   }{2 \alpha \omega \varepsilon (p_i+p_j) } & \mbox{ else. }
\end{array}
\right .     
\end{equation}

As already mentioned in Section \ref{sec_intro}, any manufacturer can decide not to operate.
Now say  manufacturer (say $M_j$) decides not to operate. Its entire  customer-base $\dbar$  along with $\alpha p_i$  fraction of   customer-base  of  manufacturer $M_i$  lookout for opportunities, again only a fraction $\varepsilon(\dbar+ \alpha p_i)$ of them are NDLP customers based on essentialness. Under this condition, there are no options for the NDLP customers, hence they all resort to $M_i$. In all, we  have the   demand model as   in \eqref{eqn_demand_overall} given at the top of the  page, for manufacturer $M_i$ using  $\mu_i^*$  of  \eqref{Eqn_NDLP_split}.

Again using \eqref{Eqn_NDLP_split}, if the customers are highly price-sensitive and less sensitive to the QoS, more precisely with $\omega = 0$, we will have the following demand function:
\begin{eqnarray}
    D_\sMi &=&  \left \{
    \begin{array}{lll}     
     \left (\dbar (1+\varepsilon)- \alpha (1-\varepsilon) p_i  \right )^+ 
    &  \mbox{if } p_j = n_o \\
    \left (\dbar - \alpha p_i + \varepsilon \alpha \mu^*_i (p_i + p_j) \right )^+      &   \mbox{else. } 
    \end{array}
    \right . 
    \label{demand_function} \nonumber \\
    \mbox{ where } && \mu^*_{i} = \frac{\indc{p_{i} \leq  p_j}}{\left| \arg\min_m { p_m}\right|} .
\end{eqnarray}

{\bf Remarks on the new demand model:}
1) 
In studies like  \cite{Wadhwa,Li, Zheng}, authors consider  a linear demand model, where customers  dissatisfied with their manufacturer  (like NDLP customers) switch to other manufacturers.  Our model differs significantly from such models in two aspects: i) the switching is a strategic decision, and not that of blindly alternating to other option (by considering mean-field game); and ii) the choice can also depend upon the QoS. It is more realistic to model that the exploring customers explore all possible choices in a strategic manner, rather than making another blind choice.

2) Our demand model explicitly accounts for both loyal and non-loyal customer segments, whose sizes themselves depend upon the prices quoted, allowing for a more realistic representation of the price-sensitive and competitive market. 

3)  The  extension of market model to any finite number of manufacturers is straight forward in our framework, due to strategic choices by NDLP customers and the eventual segmentation driven by the mean-field equilibrium. 

\subsection{ Interaction Among Agents}\label{sub_sec_inter}
\ignore{
We consider an SC with a supplier in the upper echelon and two manufacturers in the lower echelon.  
The manufacturers at the lower echelon are the followers and the supplier at the upper echelon is the leader. The supplier is responsible for the procurement of raw material to the manufacturers while the manufacturers produce the finished product and sell it to the customers as in any generic supply chain setting. We additionally consider the manufacturers to be symmetric ( they have same market potential,price sensitivity and costs)}

Manufacturer $m$ ($m$ can be $i$ or $j$)  in the lower echelon can choose to operate by quoting  a price $p_m$  for the final product   or choose   not to operate; the latter action represented by $n_o$ is chosen when the market conditions are not favorable.
Similarly the supplier can choose to quote a price $q$ to the manufacturers or not-to-operate. Thus the action of the  manufacturer $m$  is given by $p_m \in \P \cup \{n_o\}$ 
 and that of the supplier is given by $q \in  \Q \cup \{n_o\}$, for some price sets $\P$ and $\Q$ respectively.
%
The \manus incur per-unit production cost $C_\sM$ and a fixed operating cost $O_\sM$; we are studying a symmetric system in this paper and hence the values are the same for both of 
them. 
The supplier incurs a per-unit raw material procurement cost $C_\sS$ and a fixed operating cost $O_\sS$.

Let $\p = (p_i, p_j)$ and $\ba = (\p, q)$  respectively represent the joint actions of the two \manus   and  all the agents respectively. 
Any agent derives zero utility, if it chooses action  $n_o$.  Otherwise, the utility of the agent is the revenue generated, which equals the product of the demand attracted by that agent and the per-unit profit earned.  Thus the utility of the manufacturer $m$ is given by (with $\Cmq = C_\sM + q$):
\begin{equation}\label{eqn_util_manu}
U_m({\bf{a}}) = \left(D_{\sM_m}(\bp)(p_m- \Cmq )\indc{q \ne n_o} - O_\sM\right)\indc{p_m \ne n_o}
\end{equation}
where $D_{\sM_m}(\cdot)$ is defined in \eqref{demand_function}
and
the utility of the supplier is given by:
\begin{equation}\label{eqn_util_supplier}
U_\sS(\ba) = \left(\sum_{m}D_{\sM_m}(\bp)\indc{p_m \ne n_o}(q- C_\sS) - O_\sS\right)\indc{q\ne n_o}. 
\end{equation}
Basically the response  of the manufacturers (in addition to the market reputation and other factors) will determine the market segmented to each manufacturer (as in \eqref{demand_function}) and   will also   percolate to the supplier in the upper echelon. Further, the utilities and the actions  of the \manus depend upon  the price $q$ quoted by the supplier. Such interactions have to be studied using appropriate game theoretic framework.

In this paper, we consider a study where the supplier first chooses the price. Thus we
 have a Stackelberg game with the supplier in the upper echelon as the leader and the manufacturers in the lower echelon as the followers.
 There is a 
 non-cooperative strategic form game at the lower level among the manufacturers, 
 for each price $q$ quoted by the supplier. 
 The outcome $\ba^{*} = (q^*, \p^*(q^*) )$ of such a game  is called Stackelberg  equilibrium (SBE), where:
\begin{eqnarray}\label{eqn_sbe}
    q^{*} \in  \arg\max_{q} U_\sS(q,\bp^{*}(q))  
\end{eqnarray}
and $\bp^*(q)$ is a Nash Equilibrium (NE)\footnote{One may have multiple NEs for the lower game and we then choose an appropriate focal NE or Schelling point (\cite{Schelling})} of the lower level game $\G(q)$ (for each $q$), which in turn is given by:
\begin{eqnarray*}
p_m^{*}(q) = \arg\max_{p_m} U_{m}(\bp;q) \mbox{ for each }  m \in \{i, j\}.
\end{eqnarray*}
We now move on to the analysis of the game.

\section{Analysis}
We start with analyzing the non-cooperative strategic form game  $\G(q)$  among the manufacturers in the lower level, for any given price $q$ of the supplier. We will observe that one can have more than one NE, and so discuss the focal NE (see \cite{Schelling}) and then derive the SBE numerically. 

\subsection{Game among  Manufacturers}
At the lower level, for any given strategy $q$ of the leader or the supplier, the two 
manufacturers are engaged in a \textit{strategic-form non-cooperative game} denoted by
$\G(q)$  with common  strategy set $S = \P \cup \{n_o\}$
and  with utilities   as  defined in \eqref{eqn_util_manu}.
We now analyze this game for each $q \ne n_o$, i.e., when the supplier operates.
In this section, we will first show that this game has 
solutions based on the range of $q$ and then identify the solution for each regime.  We begin with delving into more details of the utility function \eqref{eqn_util_manu}, which facilitates identifying the regimes.

Say the manufacturers quote prices $(p_i, p_j)$ and say $p_i < p_j$. From \eqref{demand_function} and \eqref{eqn_util_manu},
 the manufacturer $M_i$
 attracts the entire fraction of NLDP customers $\alpha \varepsilon (p_i+p_j)$ 
 while $M_j$ gets a smaller share (only a fraction of its loyal customer-base). Hence their respective utility functions   (with  $p_i < p_j)$ are given by:

 \vspace{-3mm}
 {\small\begin{eqnarray}
   U_i (p_i, p_j)\hspace{-2mm} &=& \hspace{-2mm}W_1 (p_i, p_j) \mbox{ and } U_j (p_i, p_j) = W_2 (p_j)   \mbox{, where, } \nonumber\\
  W_1 (p_i,p_j)\hspace{-2mm} &:=&  \hspace{-2mm}
\left(\dbar - \alpha  p_i(1-\varepsilon) + \varepsilon\alpha  p_j\right)^+ \left(p_i- \Cmq\right)  -O_\sM,  \hspace{4mm} \label{Eqn_W1} \\
W_2 (p_j)\hspace{-2mm}  &:=& \hspace{-2mm}\left(\dbar - \alpha p_j \right)^+ \left(p_j-\Cmq\right)  - O_\sM \label{Eqn_W2}.
 \end{eqnarray}}
 When the manufacturers quote prices $(p_i, p_j)$ with $p_i = p_j$, the fraction of NLDP customers get divided equally among  the two manufacturers. Hence from \eqref{demand_function} and \eqref{eqn_util_manu}, their utility functions (with $p_i=p_j$) are given by:
\begin{eqnarray}
  U_i (p_i,p_j) &=& W_3 (p_i) \mbox{ and } U_j ( p_i,p_j) = W_3 (p_j)   \mbox{, where, } \nonumber\\   
  W_3(p_i) &:=& \hspace{-2mm} \left(\dbar - \alpha p_i(1-\varepsilon)  \right)^+ \left(p_i-\Cmq \right)  - O_\sM. \label{Eqn_W3}
\end{eqnarray}
When one of the manufacturers (say $M_j$) chooses the action $n_o$ and $M_i$ quotes price $p_i$, from \eqref{demand_function} and \eqref{eqn_util_manu}, the utility functions of the manufacturers   are given by:
\begin{eqnarray}
    U_i (p_i,p_j) \hspace{-2mm} &\hspace{-2mm} =\hspace{-2mm} & \hspace{-2mm} W_4 (p_i) \mbox{ and } U_j (p_i, p_j) = 0   \mbox{, where, } \nonumber\\
     \hspace{-0.5cm} W_4(p_i) \hspace{-2mm} &\hspace{-2mm} =\hspace{-2mm} & \hspace{-2mm} \left(\dbar(1+\varepsilon) - \alpha p_i(1-\varepsilon) \right)^+\left(p_i- \Cmq \right) - O_\sM . \nonumber 
     \\ \label{eqn_W_4}
\end{eqnarray}
In all, the utility function of the  manufacturer $M_i$ based on the actions of both the manufacturers is given by:
\begin{eqnarray}
\label{eqn_util_cases}
U_i(p_i,p_j; q) &=& 
\left\{
\begin{array}{ll}
W_1 (p_i,p_j)
  & \text{ if } 
 p_i <   p_j  ,  \\
W_2 (p_i)     & \text{ if 
 } p_i >   p_j , \\
W_3 (p_i) & \text{ if } p_i = p_j, \\
W_4(p_i)  & \text{ if } p_j = n_o, \\
0         & \text{ else, if } p_i = n_o.
 \end{array}
\right. 
\end{eqnarray}
The utility function $U_j$ for \manu $j$ can be defined in a similar way.

\subsubsection*{  Discrete set of prices}

Inspired  by the study of \cite{Maskin} and taking into account the practical considerations (it is obvious that prices can not be real numbers), we consider a discrete set of prices $\P$ as the options available to the manufacturers. We in fact obtain the study for  price set $\P_\delta$ for each $\delta$, where $\delta$ represents the difference between two consequent price choices, i.e., 
\begin{eqnarray}
  \P_\delta := \{0, \delta, 2\delta, \cdots\}   . \label{Eqn_P_delta}
\end{eqnarray}
From \eqref{Eqn_W1}-\eqref{eqn_util_cases}, when one extends each of $W_l$ (for $l \in \{1,2,3, 4\}$) functions to continuous set of prices $[0, \infty)$, the relaxed functions are clearly concave (can easily check using double derivatives). 
By leveraging upon this piece-wise concavity property of the relaxed  function, in this section,  we   systematically determine the set of NEs for each pair of 
$(q,\delta)$. In particular, we  establish conditions under which symmetric or asymmetric equilibria emerge. 
We begin with identifying some important regimes based on supplier price and denomination duo $(q,\delta)$, which demarcate/differentiate in the  types (and  the number) of outcomes of  the \manu game $\G(q,\delta)$.


\subsection*{  Complete choking regime} 
From  \eqref{Eqn_W1}-\eqref{eqn_util_cases}, it is clear that the \manus may not be able to survive if the supplier quotes an exorbitantly large price $q$; both  might prefer not-operating and choose $n_o$. 

Towards understanding this choking set of supplier prices,  for the moment, consider the relaxed setting of continuous prices. 
Say one of the \manus (say $m$) decides to not  operate, i.e., $p_m = n_o$;  observe from \eqref{eqn_util_cases}  that in monopoly setting $W_4$ determines the utility of the operating  manufacturer.  Hence, the operating \manu is choked to not operate if 
$
\max_{p} W_4(p) < 0.
$
Let $\bqm$ be the supplier price (in relaxed domain $[0, \infty)$) beyond which even  a lone operating \manu  (monopoly setting) can not sustain:

\vspace{-4mm}
{\small\begin{align}
   \bqm &:=  \frac{\dbar(1+\varepsilon) - \alpha(1-\varepsilon)C_\sM- 2\sqrt{\alpha(1-\varepsilon) O_\sM}}{\alpha(1-\varepsilon)}, \label{eq_bqd} \mbox{ with, }  \\ 
   W_4^{*} &:= \max_{p \ge 0} W_4(p)  = \left \{ \hspace{-2mm}
   \begin{array}{ll}
   \frac{\left(\dbar(1+\varepsilon) - \alpha(1-\varepsilon)\Cmq\right)^2}{4\alpha(1-\varepsilon)} - O_\sM,  &\hspace{-0.3cm}  \mbox{if } q \le \bqm, \\
   0,   & \mbox{else.}
   \end{array} \right.  \label{eq_W_4_star} 
\end{align}}%
(by concavity arguments as applied to \eqref{eqn_W_4} and in relaxed setting). 
Thus the 'relaxed' $W_4^{*}$ depends upon $q$ and actually equals zero, when $q \ge \bqm$.

 We refer the regime of supplier prices  where even a lone \manu prefers not to operate  \textit{as complete choking regime.} 
With discrete set of prices $\P_\delta$, the complete chocking regime is given by the following:
\begin{eqnarray}
\S_{cc,\delta} :=\left  \{ q  : W_{4,\delta}^* < 0 \right \}, \mbox{ where, } 
    W_{4,\delta}^*(\delta) = \max_{l\delta \in \P_\delta }  W_4 (l \delta) \label{eqn_scc}. 
\end{eqnarray}
Observe the above regime includes all supplier prices   $q$  beyond \( \bqm \), i.e.,  $ [\bqm, \infty) \subset \S_{cc, \delta}$.

{\bf Not Operating ($n_o, n_o$) is the only NE in $\S_{cc,\delta}$:}
 For any  $q \in \S_{cc,\delta}$, the complete choking regime, it is immediate that $(n_o, n_o)$ is a NE, as  
$$
U_m (n_o, n_o) = 0 > U_m (l\delta, n_o) \mbox{ for all } l\delta  \in \P_\delta, 
$$
and hence $n_o \in \BR_m(n_o)$. In fact in similar lines,  by definition of $\S_{cc, \delta}$ and from \eqref{eqn_W_4}-\eqref{eqn_util_cases},   for any $p_{-m}$, we have $\BR_m(p_{-m}) = \{n_o\}$; thus $(n_o, n_o)$ is the only pure strategy NE here. 

Thus when the supplier quotes an exorbitantly large price, none of the \manus would operate.  We now shift the focus to operating NEs, for that  now consider $q \notin \S_{cc,\delta}$.
\textit{We refer a pure strategy NE  $(p_i^*, p_j^*)$ as an operating NE, if $p_m^* \ne n_o$ for both $m.$}
We will require  the following assumption (which is assumed throughout), which ensures the market has sufficient potential for `survival' (see \cite{Wadhwa}):\\
 \noindent{\bf A}.1 The market potential is sufficiently large, \\
$$
\dbar>\alpha (C_\sS + C_\sM) + 2\sqrt{\alpha O_M}. 
$$

\subsection*{Duopoly regime and operating NEs}

If  the supplier quotes a smaller price,  we will have \textit{duopoly regime.}  From \eqref{eqn_util_cases}, $ W_2$ is the utility of \manu setting strictly higher price for final product. Thus the following can be defined as the \textit{duopoly regime}:
\begin{eqnarray*}
\S_{dp,\delta} = \left  \{ q  : W_{2,\delta}^* \ge  0 \right \},  \mbox{ where, }   
    W_{2,\delta}^*(\delta) := \max_{l\delta \in \P_\delta  }  W_2 (l \delta)
\end{eqnarray*}

We begin with investigating the conditions for existence of symmetric operating NEs -- basically when $(l^*\delta, l^*\delta)$ is a NE of $\G(q,\delta)$ for some common $l^*$ and $(q, \delta)$.

It is clear from \eqref{eqn_util_cases}  that for such an existence, we  \textit{further require that $W_3(p) \ge 0$  for at  least one $(p, p)$}, which directly implies  $q \le \bqs $ (see \eqref{Eqn_W3} and derived using relaxed setting), 
 where,   
\begin{eqnarray}\label{eqn_q_bar_s}
     \bqs :=  \frac{\dbar  - \alpha(1-\varepsilon)C_\sM- 2\sqrt{\alpha(1-\varepsilon) O_\sM}}{\alpha(1-\varepsilon)}.\label{eq_bqs}
\end{eqnarray}
 We now have the first main result; all the proofs are provided in   Appendix A.

\begin{thm}\label{thm_sym_NE}
{\bf [Operating symmetric  NEs]} Consider any $q  \notin \S_{cc, \delta}$ and $q \le \bqs$ . 
 Let $s_\delta(q) \le  e_\delta (q) $,  for the coefficients defined in \eqref{Eqn_coefficients}, of Appendix A.  
 
 If  $l$ is a positive integer with $l\delta \in  [s_\delta(q), e_\delta(q)]$,
  then 
    $(l\delta,l\delta)$   is a symmetric pure   NE of the    \manu game $\G(q,\delta)$. 
\end{thm}

The above theorem identifies the conditions in which both the \manus operate and in fact derive equal non-zero profits. The supplier price should be conducive enough for such a symmetric survival; they derive equal profit margins at the NEs of Theorem \ref{thm_sym_NE}. 
Also observe here the denomination factor $\delta$ plays a significant role in the existence of such NEs.

It might be possible that  both the \manus  survive and operate profitably, but  at  different profit margins. 
 We immediately investigate such a possibility.

\subsection*{Duopoly regime and asymmetric operating NEs}

Before proceeding to Theorem \ref{thm_asym_NE_befor_q_bar} which establishes the existence of asymmetric NE,  we would require some important definitions.

As seen before, the piecewise (relevant) functions are concave in the relaxed setting and  would have an unique optimizer in any bounded  or unbounded regime. We would actually require the corresponding optimizers in the discrete  set, $\P_\delta$.  
Hence we define  $\Delta(f,\delta) := \arg \max_{p \in \P_\delta} f(p)$ as the operator that assigns the optimizer of function,   to  any  pair $(f,\delta)$, where  $f$ is a function  with unique optimizer. 
For the sake of mathematical tractability, we consider the tie breaker to be the  floor value, $\lfloor{. \rfloor}$ (the right side optimizers can drift to other regimes):

\vspace{-4mm}
{\small
\begin{eqnarray*}
 \Delta(f,\delta) = \left \{
    \begin{array}{lll}     
    \lfloor{ \frac{\arg\max f}{\delta}\rfloor} \delta
    &  \mbox{if }  f(\lfloor{ \frac{\arg\max f}{\delta}\rfloor} \delta) \ge f(\lceil{ \frac{\arg\max f}{\delta}\rceil} \delta)\\
            \lceil{ \frac{\arg\max f}{\delta}\rceil} \delta
        & \mbox{else. } 
    \end{array}
    \right .  
\end{eqnarray*}}
Using the above define:
\begin{eqnarray}
\label{Eqn_lbard}
 \ld \delta  = \Delta(W_2,\delta).
\ignore{
    \ld \delta  \in \arg \max_{l\delta \in \P_\delta \setminus \{n_o\} } W_2 (l\delta).}  
\end{eqnarray}
By definition, $\ld \delta \in \P_\delta$  and
 satisfies:
 \begin{eqnarray}\label{eqn_w_2_star_delta}
   W_{2,\delta}^* = W_2 (\ld \delta). 
 \end{eqnarray} 
We now have the next result about asymmetric operating NEs.
\begin{thm}\label{thm_asym_NE_befor_q_bar}
{\bf [Asymmetric NEs]}  
For supplier prices in the duopoly regime,
i.e.,   for $q \in \S_{dp, \delta}$, ,  one can have asymmetric $(\ld\delta, (\ld-1)\delta)$ and $((\ld-1)\delta, \ld\delta)$ as NEs, under certain conditions provided in Appendix A.
For 
any   $q \notin  \S_{dp, \delta}$, one  will not have asymmetric NE.  
 \end{thm}
 Interestingly both the \manus can survive and operate profitably, even by deriving different amounts of   profits. Such an asymmetric survival is possible only because of discrete set of prices. 
 Observe here that in the duopoly regime,  one can also symmetric NEs of Theorem \ref{thm_sym_NE}, where the \manus operate at equal profit margins.

\subsection*{No equilibria for smaller denomination factor $\delta$}
Finally we show that the game $\G(q, \delta)$ will have no operating NE for small enough $\delta.$ 
\begin{thm}\label{thm_no_NE} {\bf [No operating NE for small $\delta$]} For any supplier price $q$, 
  there exists a $\underline{\delta} >0$, such that for any $\delta \le \underline{\delta}$,   there exists no operating NE.
    \end{thm}

{\bf Remarks:} Our 
problem setting
differs from the well known Bertrand duopoly model in two main  aspects.  As already mentioned, our demand  model \eqref{demand_function} considers loyal and strategic customers, with price sensitive loyalties. Secondly, the utility  models in \eqref{eqn_util_manu}-\eqref{eqn_util_supplier} incorporate yet another important practical consideration, that of operating costs. This results in significantly different outcomes as proved in Theorems~\ref{thm_sym_NE}-\ref{thm_no_NE}. 
\begin{itemize}
    \item[i)] The marginal prices can no longer form the NE; due to operating costs we in fact  have $(n_o, n_o)$  as the NE, indicating \manus in reality would not operate at marginal prices. 

    \item [ii)] The consideration of discrete prices as in \cite{Maskin} did not  result in Edgeworth Cycles (it is easy to observe that
    every NE becomes an 'optimal policy'  for the dynamic game of  \cite{Maskin}   in the following sense --- at  the corresponding optimal policy, the players choose  the NE prices at every time slot) (see Figure \ref{fig:NE_Convg})   . Such an outcome is resultant of three important factors-- operating costs, a mixture of strategic and loyal customers and discrete set of prices; only the  last factor is common with \cite{Maskin}. 

    \item [iii)] The NEs disappear once the denomination  factors reduce beyond a limit. It might be practical to consider discrete set of prices, however our work is still limited as it does not consider a general discrete set.  It would be interesting to work with a more general set of prices and elaborate   and realistic demand model  \eqref{eqn_demand_overall} as proposed in this paper (which in fact also incorporates QoS considerations)  and  such a study  will be a part of  our future research. 
\end{itemize}

\section{Numerical analysis and  Supplier's perspective}

We now provide a numerical study of the NEs obtained in Theorems \ref{thm_sym_NE}-\ref{thm_no_NE}. We also discuss the optimal choice of the supplier, given the multitude of the solutions (or no solutions) at the lower level. We set $\dbar = 8, C_\sM =O_\sM = 2$,  $C_\sS =O_\sS =0.01$,  $\alpha = 0.5$ and $\varepsilon = 0.8$, in all the examples studied below.

\subsection*{Variation Of NEs with Supplier's Price $q$ } 
\ignore{
We vary parameters $\alpha, \varepsilon$ and $\delta$ to understand the variations in the number of NEs and tabulate the same in Table \ref{table_alpha}. When price sensitivity parameter $\alpha$ of the manufacturers is large (see rows $1,2,3$ and $4$) the number of NE's is less as compared to the case when  the price sensitivity is less (see rows $5,6,7$ and $8$). This happens because the customers are highly sensitive to price changes, and the manufacturers have fewer viable pricing options, leading to fewer Nash Equilibria. Likewise when essentialness of the product $\varepsilon$  is high (see rows $5$ and $6$) the number of NEs is high while that  for low value of $\varepsilon$ (see rows $7$ and $8$), is smaller; this is  because the more  essential  the product is,  the more will be the affinity of the customers to buy it.
No such monotonous trend in the variation of number of NEs is seen when the denomination factor $\delta$ varies. }

We conduct a numerical study to investigate variations in the set of symmetric operating NEs as the supplier's price increases, as illustrated in Figure \ref{fig:ne}. With increase in $q$, the number of  NEs decrease and then eventually disappear. This disappearance of NEs is observed  at a much smaller supplier price when the denomination factor is small ($\delta = 0.8$); this  indicates that smaller denomination factors foster more instability in  the downstream market. In fact 
all  the NEs just disappear  for $\delta = 0.02$, as established in Theorem \ref{thm_no_NE}.

\subsection* {Supplier's Game}
The supplier as the leader sets the price at the top-level.  For some supplier prices, either there exist no NE (indicating instability) or none of the \manus prefer to operate; the supplier would naturally   avoid  quoting   such prices. For some other quoted prices,  we have multitude of NEs; it is natural to anticipate that the \manus would settle to  one among the NEs that provides the maximum utility  to each of them; the prices with asymmetric NE  are also avoided, as they can also indicate instability due to price wars. \textit{Such an NE represents the focal NE} (\cite{Schelling}).  

In view of the above reasons, we consider a numerical study of the supplier's game as in the following (see \eqref{eqn_util_supplier}):
\begin{eqnarray}\label{eqn_util_sup}
    U_\sS(q) = 2\left(\dbar - \alpha (1-\varepsilon) l^*(q) \delta \right)^{+}(q- C_\sS) - O_\sS.
\end{eqnarray}
where $q $ is restricted to those prices which result in symmetric NEs of Theorem \ref{thm_sym_NE} and then  the focal NE,
$$
l^*(q) \in \arg\max_{l : l \delta \in [s_\delta(q), e_\delta (q)]} W_3 (l\delta).
$$
Figure \ref{fig:supplier_utility}
demonstrates the  variation of total utility of the supplier with its quoted price $q$ at low and high $\delta$ values.  The vertical line in the figure represents the supplier's price threshold beyond which one of the manufacturers gets partially choked   ---  in this regime a  \manu setting  a strictly bigger price  derives negative utility and hence 
the two \manus set equal price (for the final product) at the corresponding focal NE.
Interestingly, the supplier finds it beneficial (derives highest utility) to quote such  a price (beyond the threshold) taking  an advantage of it's leadership in the SC.

\subsection*{Best response dynamics}
\begin{figure*}[t] 
\vspace{-1.3cm}
\centering

\begin{minipage}{0.38\textwidth} 
    \centering
    \includegraphics[width=0.85\linewidth]{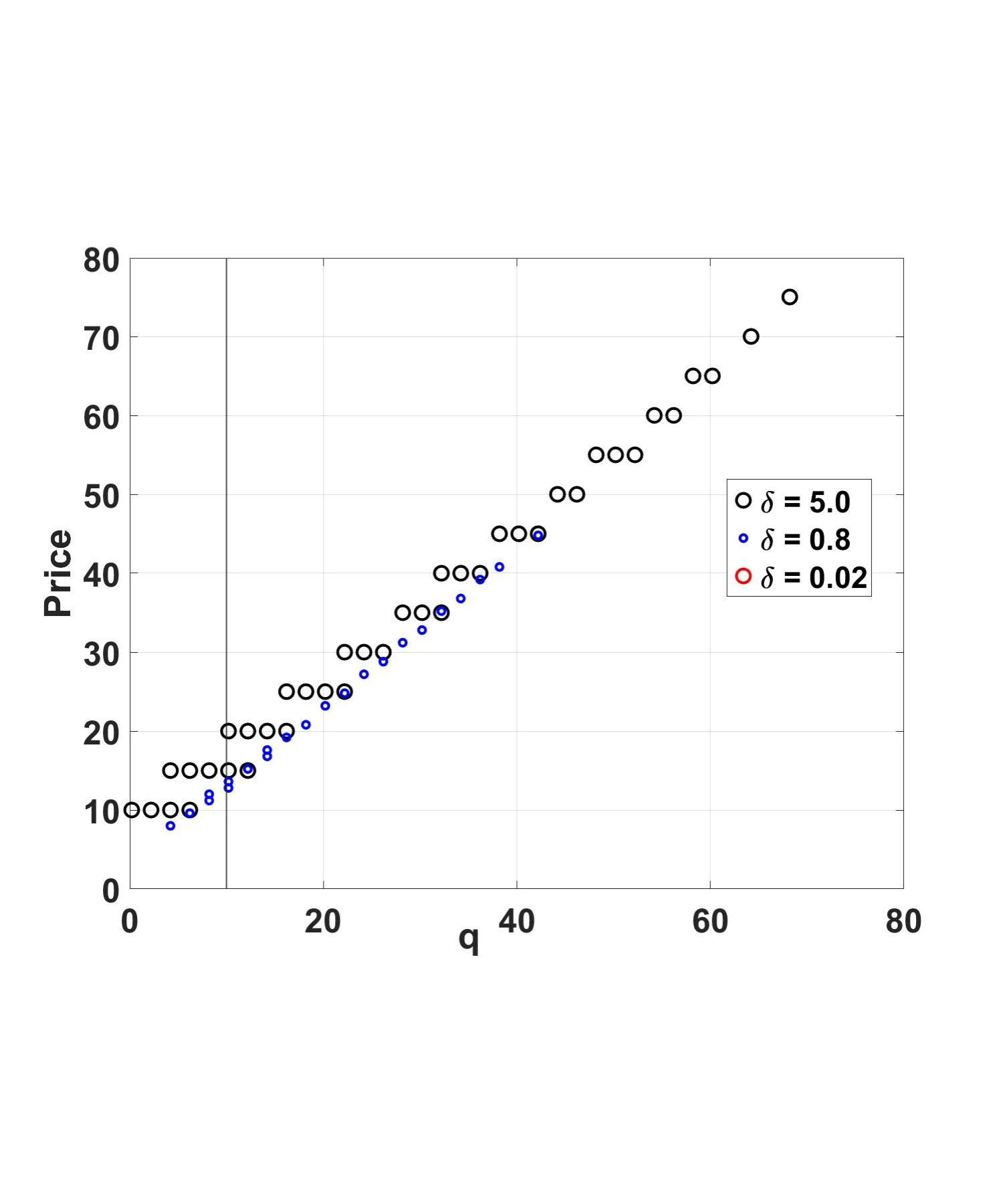}
    \vspace{-12mm}
    \caption {The variations in symmetric NE sets with supplier prices and price denominations} 
    \label{fig:ne}
\end{minipage}
\hspace{1cm} 
\begin{minipage}{0.38\textwidth}
    \centering
    \includegraphics[width=0.85\linewidth]{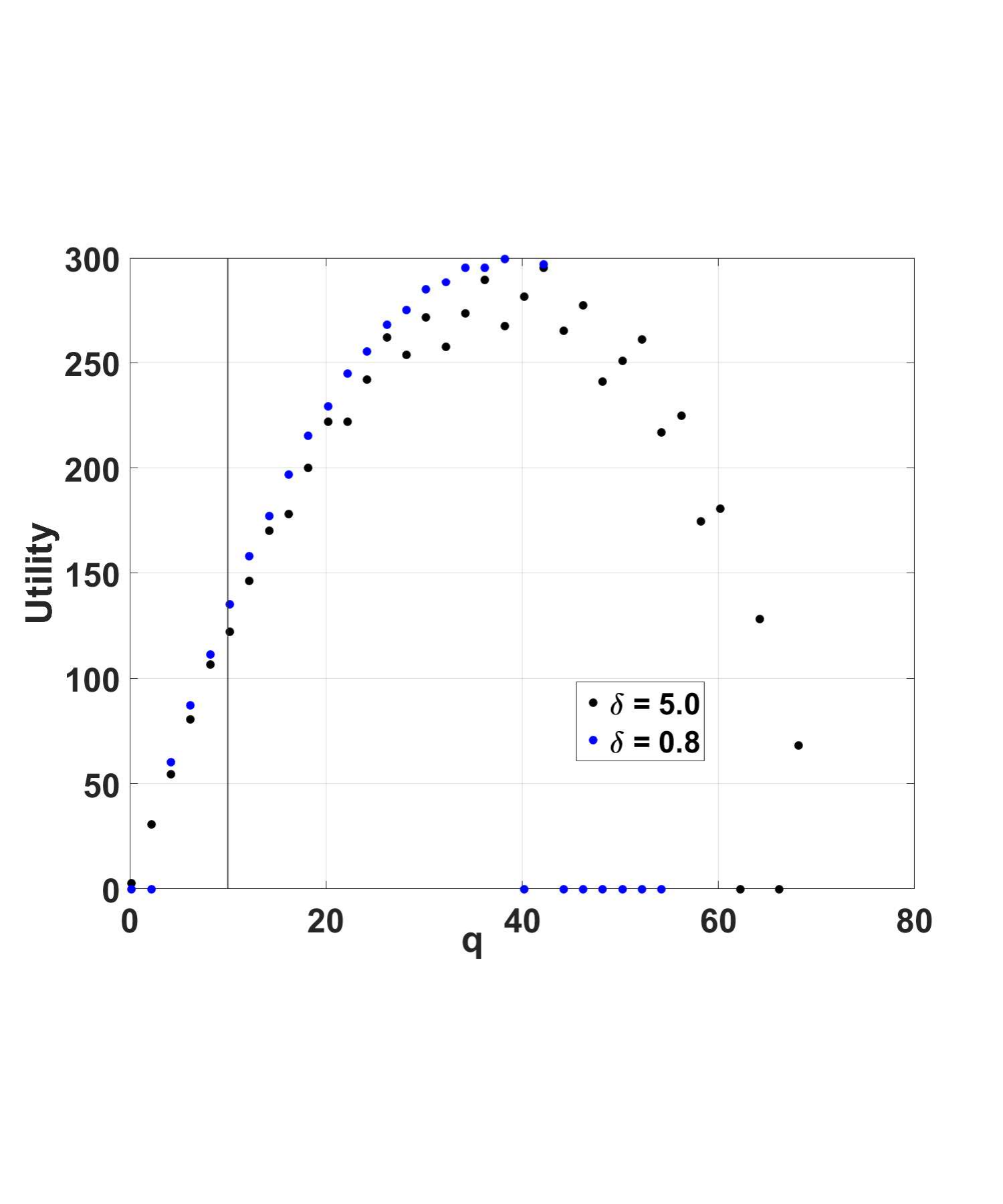}
    \vspace{-12mm}
    \caption{Utility function of the supplier at focal NE of the \manus}
    \label{fig:supplier_utility}
\end{minipage}

\vspace{-11mm} 

\begin{minipage}{0.38\textwidth} 
    \centering
    \includegraphics[width=0.9\linewidth]{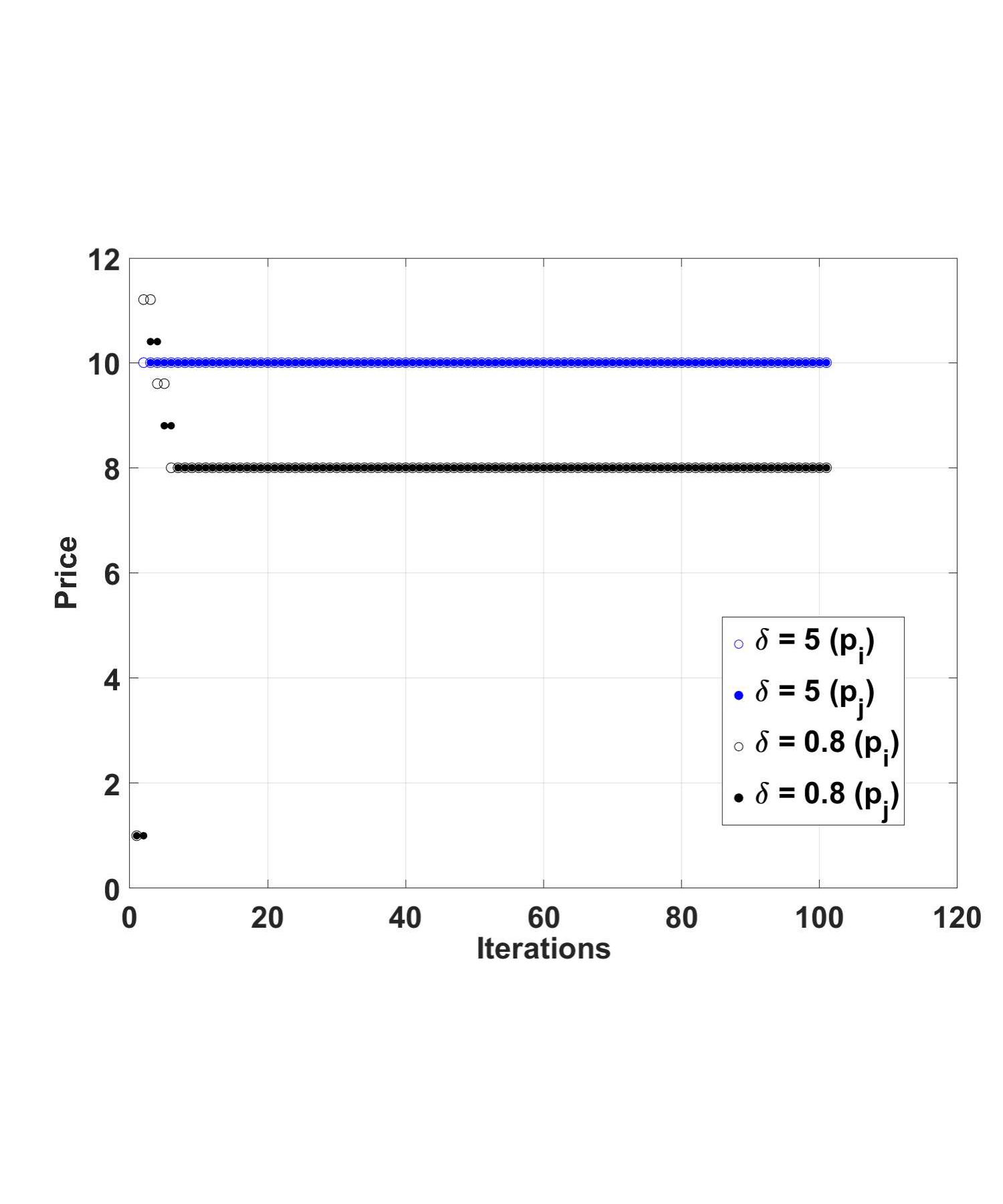}
    \vspace{-12mm}
    \caption{Convergence to an NE: for     $\delta$,  where existence of NE is proved in Theorem \ref{thm_sym_NE}.}
    \label{fig:NE_Convg}
\end{minipage}
\hspace{1cm} 
\begin{minipage}{0.38\textwidth}
    \centering
    \includegraphics[width=0.9\linewidth]{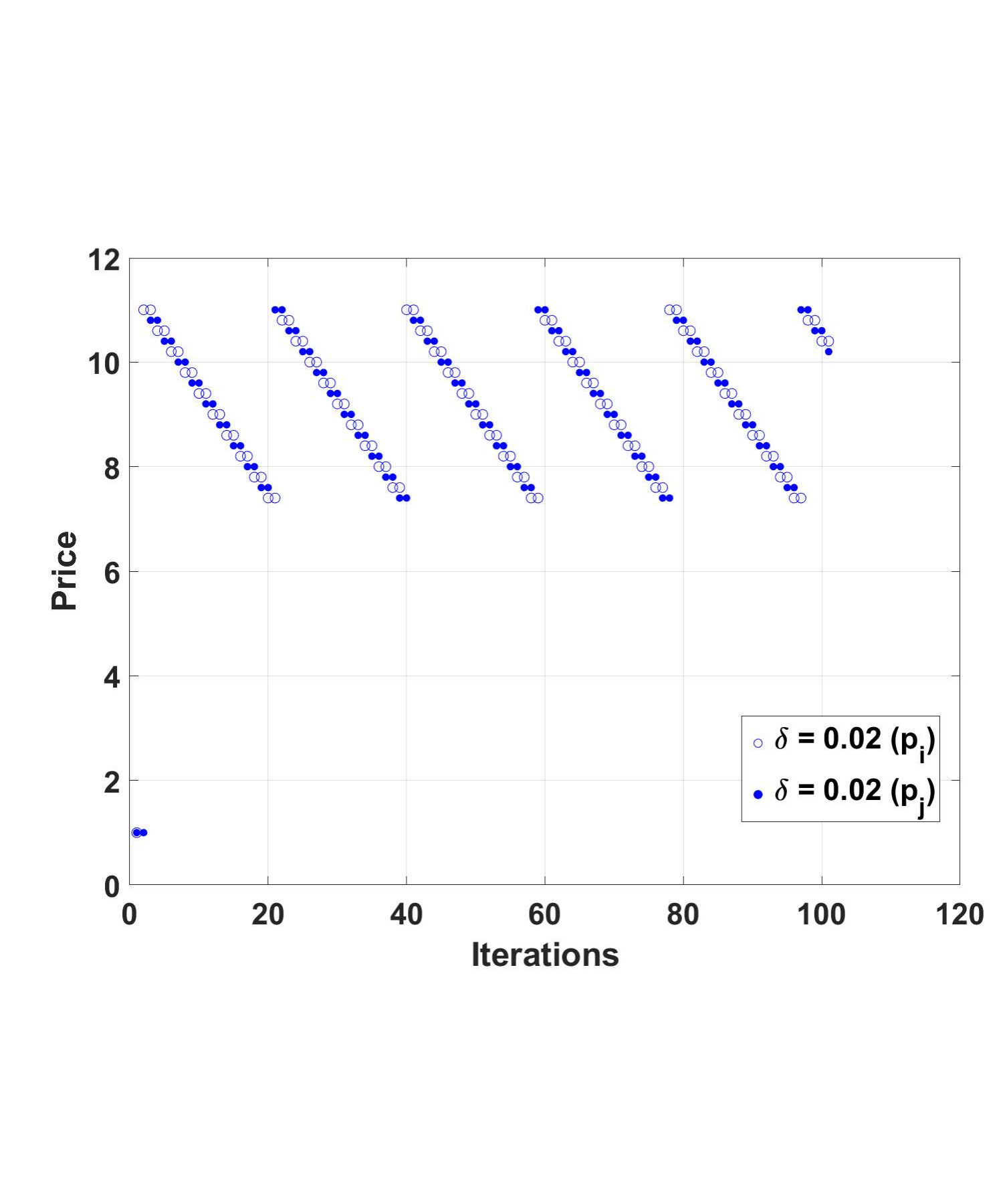}
    \vspace{-12mm}
    \caption{Convergence to Edgeworth Cycle: for small $\delta$ with no NE, as given in Theorem \ref{thm_no_NE}. }
    \label{fig:Cycle}
\end{minipage}

\label{fig:four-figures}
\end{figure*}

We conducted a numerical study of the
best response dynamics.  At every iteration $k$, one of the agents (say \manu $m$) finds a best response against strategy $p_{-m,k}$ of its opponent and replaces its strategy with that best response. The  set of parameters are same  as that mentioned in the beginning of this section and for the fixed supplier's price, $q = 4.2$. Unlike their findings of Edgeworth Cycles, the best response dynamics eventually converge to NE for large denomination factors ($\delta = 5$ and $\delta = 0.8$) as shown in Figure \ref{fig:NE_Convg}. 
Interestingly we observe convergence to Edgeworth Cycles   for  small denomination factors ($\delta = 0.02$) as seen in Figure \ref{fig:Cycle}; from Theorem \ref{thm_no_NE}, we do not have NE for such~$\delta$. 

For the dynamic game  of \cite{Maskin}, Edgeworth cycle is a Markov perfect equilibrium, so is the resultant of our best response dynamics for small $\delta$ (in Figure \ref{fig:Cycle}). However for larger $\delta$, our dynamics converges to one of the NEs, and one can prove this constant curve to  be the equilibrium even for the dynamic game  of \cite{Maskin} for such $\delta$; as already mentioned this contrast arises due to consideration of realistic aspects like  loyal-strategic customers and operating costs in our  models.

\section{ Conclusions}

The  well-known Bertrand duopoly illustrates the existence of NE at marginal prices.
Maskin and Tirole argued against the Bertrand equilibrium by observing that agents would not settle at marginal prices. Considering  
 discrete set of prices 
 and a restricted dynamic game (depends only on previous slot actions);
they showed that Edgeworth Cycles form Markov Perfect Equilibria.  
 and disputed the marginal price equilibrium. 

 By considering a more realistic setting with non-zero operating costs, loyal-strategic customer tendencies and discrete price settings, we illustrate that we neither have marginal price based equilibria nor do we have price cycles as solutions. We establish the existence of equilibria with positive profit margins (unless the supplier quotes exorbitantly high). 

 This study opens up a lot of new questions, which require further investigation.   
   Will the NE still exist when a more general
set of prices (with all kinds of denomination factors) is considered? Will we really have cycles in such a generic setting (we observe some in numerical examples)? Further, how will the overall equilibrium
behavior of the SC agents change with asymmetric agents?

\begin{figure*}
  {\small\begin{eqnarray}
      s_\delta (q) &\hspace{-2mm} =\hspace{-2mm} &
      \max \left\{ \frac{  \lambda_s- \sqrt{ {\underline \lambda}_s^2 - 4\alpha (1-\varepsilon) O_\sM} }{2\alpha (1-\varepsilon)},    \frac{\lambda_s - \sqrt{\lambda_s^2 + 4 \alpha (1-\varepsilon) \varrho_s}}{2    \alpha (1-\varepsilon)}  \indc{q \in \S_{dp,\delta}}  \right \}  =  \frac{\lambda_s - \sqrt{\lambda_s^2 + 4 \alpha (1-\varepsilon) \varrho_s}}{2    \alpha (1-\varepsilon)}     ,     \mbox{ with }    \label{eqn_s_q} \\
      \lambda_s  &\hspace{-2mm} =\hspace{-2mm} &  \dbar + \alpha(1-\varepsilon)\Cmq, \ \ \varrho_s  = 
       -\dbar\Cmq - \left (W_{2,\delta}^{*} \right )^+- O_\sM,   \mbox{ and } {\underline \lambda}_s = \dbar - \alpha (1-\varepsilon)\Cmq .  \nonumber \\      e_\delta (q) &\hspace{-2mm}=\hspace{-2mm}&   \min\left \{ 
      \frac{\lambda_s + \sqrt{\lambda_s^2 + 4 \alpha (1-\varepsilon) \varrho_s}}{2    \alpha (1-\varepsilon)}, \tilde{e} 
      \right \} \indc{q \in \S_{dp,\delta}} + \tilde {e} \indc{q  \in \S_{pc,\delta} },
      {\tilde e}  =  \min\left \{     \frac{\lambda_e + \sqrt{\lambda_e^2 + 4 \alpha \varepsilon \varrho_e}}{2    \alpha \varepsilon} , \frac{ \lambda_s + \sqrt{ {\underline \lambda}_s^2 - 4\alpha (1-\varepsilon) O_\sM} }{2\alpha (1-\varepsilon)}     \right \} \nonumber  \\
      \lambda_e &\hspace{-2mm}=\hspace{-2mm}&  \alpha \Cmq \varepsilon + \alpha \delta (3 \varepsilon - 2), \mbox{ and, }     \varrho_e  = \delta \alpha (1 - \varepsilon)(\Cmq + \delta) +\dbar \delta 
       . \label{Eqn_coefficients}
  \end{eqnarray}}
 \ignore{ \begin{equation}
  l\delta \in   \left [  \frac{ \lambda_s- \sqrt{ {\underline \lambda}_s^2 - 4\alpha (1-\varepsilon) O_\sM} }{2\alpha (1-\varepsilon)},  \frac{ \lambda_s + \sqrt{ {\underline \lambda}_s^2 - 4\alpha (1-\varepsilon) O_\sM} }{2\alpha (1-\varepsilon)} \right ] \hspace{2mm} \label{Eqn_tight}   
  \end{equation}
with 
$
 {\underline \lambda}_s = \dbar - \alpha (1-\varepsilon) \Cmq. 
$}
\hrule 
  \end{figure*}
\section*{ Appendix A}
{\bf Proof of Theorem \ref{thm_sym_NE}:} 
To begin with we obtain some initial conditions that lead towards an operating NE.
A pair $(l\delta, l\delta)$ can become an operating NE   only if 
$
W_3 (l\delta) \ge  0$  which requires  $q  \notin \S_{cc.\delta} $ and the following (using the concavity of the relaxed~$W_3$ function, using \eqref{eqn_s_q}), 

\vspace{-3mm}
{\small\begin{eqnarray*}
  \  l\delta \in \left (\Cmq, \frac{\dbar}{\alpha (1-\varepsilon)} \right ) 
  \end{eqnarray*}
  \begin{equation}
  l\delta \in   \left [  \frac{ \lambda_s - \sqrt{ {\underline \lambda}_s^2 - 4\alpha (1-\varepsilon) O_\sM  } }{2\alpha (1-\varepsilon)},  \frac{ \lambda_s + \sqrt{ {\underline \lambda}_s^2 - 4\alpha (1-\varepsilon)  O_\sM  } }{2\alpha (1-\varepsilon)} \right ] \hspace{2mm} \label{Eqn_tight}   
  \end{equation}}%
  By \eqref{eqn_q_bar_s} the above interval is well defined. 
Clearly,

\vspace{-3mm}
{\small\begin{eqnarray*}
\frac{ \lambda_s - \sqrt{ {\underline \lambda}_s^2 - 4\alpha (1-\varepsilon) O_\sM  } }{2\alpha (1-\varepsilon)} > \Cmq \mbox{ and }  \\ 
 \frac{ \lambda_s + \sqrt{ {\underline \lambda}_s^2 - 4\alpha (1-\varepsilon)  O_\sM  } }{2\alpha (1-\varepsilon)} 
<  \frac{\dbar}{\alpha (1-\varepsilon)},
  \end{eqnarray*}
}
and so 
we only need~\eqref{Eqn_tight}.
\ignore{
We first show that  $n_o$ is not a part of an NE with $q  \in \S_\delta$.
By definition, clearly:
$$
\BR_m(n_o) \subset \{ l\delta : W_4 (l\delta) > 0\},
$$
which is non-empty for $q \in \S_\delta$ 
and hence $(n_o, n_o)$ is not a NE.  
Consider any  $l^*\delta \in  \arg \max_{l\delta \in \P_\delta\}}  W_4 (l\delta)$. Clearly $n_o \notin \BR_m(p') $ as  $W_3(p') > 0.$

{\color{red} 
Explain why $(n_o, n_o)$ or $(lq, n_o)$  is not a NE for any $q  < \bqm$.  Just show $\BR_m(n_o)$ is monopoly price $p^*(q)$ and  $n_o \notin \BR_m (p^*(q) ) $; this is true for all $q \le \bqm$. 

Also observe one can never have NE at which agents derive negative utility, because $n_o$  is  strictly  better than any action that yields negative utility.  Thus in the following we only consider pair of prices for which the utilities of both the agents are positive. 
}
}
Consider any agent $m \in \{i, j\}$.
   We begin with finding   its best responses   against all  $p_{-m} \ne n_o$.
Define the following set:
\begin{eqnarray*}
\S_{pc,\delta} :=\left  \{ q  : W_{2,\delta}^* < 0 \right \}
\end{eqnarray*}
This set can be interpreted as the  \textit{partial choking regime} in which one manufacturer stops operating and a lone manufacturer can operate.

Against any $l\delta \in \P_\delta$ of opponent (see \eqref{Eqn_P_delta}), clearly (by   concavity of the relaxed  functions $W_1, W_2, W_3$ that build the utility function, see \eqref{eqn_util_cases})  the best response $\BR_m(l\delta)$ of \manu $m$ is among 
\begin{eqnarray}
    \BR_m(l\delta) &\subset& \{(l-1)\delta, l\delta, \ld\delta, n_o\} \mbox{ if } q \in \S_{dp,\delta}  \nonumber \\
       \BR_m(l\delta) &\subset & \{(l-1)\delta, l\delta,   n_o\} \mbox{ if } q \in \S_{pc,\delta} , \label{Eqn_BR_sets}
 \end{eqnarray}
 recall by definition \eqref{Eqn_lbard},  we have $W_2^{*} = W_2(\ld\delta) \ge 0 $  and thus a chance for $\ld\delta$ being in BR set for $ q \in \S_{dp,\delta} $.

{\bf When $q \in \S_{pc,\delta}$:}
 From \eqref{Eqn_BR_sets},
one can't have an asymmetric NE in this case (as $(l+k)\delta$  with some $k \ge 1$  is never in BR set against $l\delta$).

Now      symmetric pair   $(l\delta,  \ l\delta)$ for some $l$ is   an operating NE,   iff:    i) $l\delta \in \BR_m(l\delta)$ for each $m$; and ii) $l$ satisfies \eqref{Eqn_tight}. 
Now  using \eqref{Eqn_tight}-\eqref{Eqn_BR_sets} such a pair is  an operating NE   iff $v(l)  \ge  0$  where,
\begin{eqnarray}\label{eqn_vl}
    v(l) &:=& U_m (l\delta, l\delta)  -  U_m ((l-1)\delta , l\delta)  \\
    &=& W_3 (l \delta) - W_1((l-1)\delta , l\delta), \nonumber 
\end{eqnarray}
By \eqref{Eqn_tight}   $W_3(l\delta) \ge 0$, 
and now 
  from equations \eqref{Eqn_W1}, \eqref{Eqn_W3} and \eqref{eqn_util_cases},
 $v(l)$ can be rewritten as below, using the coefficients  defined in the hypothesis of the theorem:

\vspace{-4mm}
{\small\begin{eqnarray}
v(l)   &=& -(l\delta)^2 \alpha \varepsilon + (l\delta)  \lambda_e   + \varrho_e,    \label{Eqn_vl}
\end{eqnarray}}
The above is quadratic in $l\delta$, has two real roots with only one of them positive and $v(0) > 0$. Hence $v(l) \ge  0$ for 
\begin{eqnarray}
    l\delta \le \frac{\lambda_e + \sqrt{\lambda_e^2 + 4 \alpha \varepsilon \varrho_e}}{2  \alpha \varepsilon}, \mbox{ and } l\delta \mbox{ satisfying \eqref{Eqn_tight}.}
    \label{Eqn_conditions_for_q_gt}
\end{eqnarray}
This proves the result for  $q  \in \S_{pc,\delta}.$

{\bf When $
q \in \S_{dp,\delta}$: }
%
We begin with finding all possible symmetric operating NE $(l\delta,l\delta)$. 
Now   from \eqref{Eqn_BR_sets},  for  $l \delta$ to be in  $\BR_m(l\delta)$, we require $v(l)$ of \eqref{Eqn_vl} to be non-negative (as in previous case); we additionally require  $W_{2,\delta}^* \le  U_m(l\delta, l\delta)$. 
 Thus again using \eqref{eqn_vl}-\eqref{Eqn_vl}, the tuple $(l\delta, l\delta)$ is a symmetric operating NE when the conditions in \eqref{Eqn_conditions_for_q_gt} are satisfied and additionally when: 
$$
U_m(l\delta,l\delta) - W_{2,\delta}^{*} \ge 0.
$$
Observe $W_{2,\delta}^* = W_2 (\ld \delta) \ge 0$, thus
using \eqref{Eqn_W2} and \eqref{Eqn_W3}    the above is satisfied when (constants as in \eqref{Eqn_coefficients}: 

\begin{eqnarray*}
  -(l\delta)^2 \alpha(1- \varepsilon)+ (l\delta)  \lambda_s   + \varrho_s   \ge 0.
\end{eqnarray*}
Observe the above quadratic equation is negative at $l\delta = 0$, has real roots only when 
$$
\lambda_s^2  + 4 \alpha (1-\varepsilon) \varrho_s > 0, 
$$
which is true  as by assumption {\bf A}.1:
$$
(\dbar + \alpha \Cmq)\varepsilon (\dbar - \alpha \Cmq) + \varepsilon^2 \alpha^2 \Cmq^2 > 0. 
$$
Thus in addition to \eqref{Eqn_conditions_for_q_gt}  (apart from $l\delta > \Cmq$) we require:

\vspace{-4mm}
{\small$$
l\delta \in \left [ \frac{\lambda_s - \sqrt{\lambda_s^2 + 4 \alpha (1-\varepsilon) \varrho_s}}{2    \alpha (1-\varepsilon)} , \frac{\lambda_s + \sqrt{\lambda_s^2 + 4 \alpha (1-\varepsilon) \varrho_s}}{2    \alpha (1-\varepsilon)}  \right ]. 
$$}

This completes the proof. \eop

{\bf Proof of Theorem \ref{thm_asym_NE_befor_q_bar}:}  Consider w.l.g. that player $i$ gets higher price at NE (if there exists one such NE). 
 Clearly any $(l\delta, l'\delta)$ with $l' < l$ cannot be an NE as $l'$ can only take $l-1$ as already mentioned unless $l = \ld \delta$ . Thus $(\ld ,\ld -1)$ is a possible candidate for NE. This can only happen when the following are satisfied 
 $$
 (\ld-1) \delta \in \B_m (\ld \delta)  \mbox{ and }  \ld \delta \in \B_m ( (\ld-1) \delta) \mbox{ for any } m.
 $$
 Thus
   for satisfying $\ld \delta \in \B_m ( (\ld-1) \delta)$   we require  (see \eqref{eqn_w_2_star_delta},\eqref{eqn_util_cases}) 
\begin{eqnarray*}
W_{2,\delta}^* =  U_\sMi(\ld  , \ld - 1) &\ge& U_\sMi(\ld -2,\ld - 1)), \mbox{ and } \\
W_{2,\delta}^* = U_\sMi(\ld, \ld-1) &\ge& U_\sMi(\ld -1, \ld -1 )  
\end{eqnarray*}
and  for $(\ld-1)\delta \in \B_m (\ld \delta)  $ we require:
\begin{eqnarray*}
U_\sMi(\ld-1 , \ld) &\ge& U_\sMi (\ld,\ld))
\end{eqnarray*}
This happens when 
\begin{eqnarray}
    \ld \delta &\ge&  \frac{\lambda_e + \sqrt{\lambda_e^2 + 4 \alpha \varepsilon \varrho_e}}{2  \alpha \varepsilon},  \mbox{ and, } \\
     W_{2,\delta}^* &\ge & \min\left \{ U_\sMi(\ld -2,\ld - 1)),   U_\sMi(\ld -1, \ld -1 ) \right \}. \nonumber
    \label{Eqn_conditions_assym_ne}
\end{eqnarray}
The above provides the conditions for existence of asymmetric NE. 
\eop

{\bf Proof of Theorem \ref{thm_no_NE}:}  Observe that the limit   $\lim_{\delta\to 0} \tilde {e} =\Cmq  \le s_\delta (q) $. Observe  that even if  $\Cmq  \in \P_\delta$, it can't be a part of any operating NE as $U_m (\Cmq, p_{-m}) < 0$ irrespective of $p_{-m}$, see \eqref{eqn_util_cases}. 
Thus for small enough $\delta$ one can't have symmetric NE. 

{\bf As $\delta \to 0$,} the asymmetric NE also vanishes, 
for example consider the following, which 

\begin{eqnarray*}
    U_\sMi((\ld -1)\delta, (\ld -1 )\delta)   - W_{2, \delta}^* =  W_3 ((\ld - 1 )\delta)- W_{2, \delta}^* \\
    \stackrel{\delta \to 0}{\to} W_3 (\pup^d) - W_2^* > 0.
\end{eqnarray*}
 Thus there exists $\bar \delta$ sufficiently small for which even asymmetric NE does not exist. This completes the proof. 
 \eop

\end{document}